\documentclass[11pt]{article}

\usepackage{amsmath,amsfonts,amssymb,theorem,cite}
\usepackage{graphicx,psfrag,tabularx,epsfig}
\usepackage[latin1]{inputenc}
\usepackage{a4,t1enc}

\newcommand{\hphi}{\hat\varphi}
\newcommand{\tphi}{\tilde\varphi}
\newcommand{\hx}{\hat{x}}
\newcommand{\tx}{\tilde{x}}
\newcommand{\hq}{\hat{q}}
\newcommand{\tq}{\tilde{q}}
\newcommand{\hp}{\hat{p}}
\newcommand{\tp}{\tilde{p}}
\newcommand{\tV}{\tilde{V}}
\newcommand{\tA}{\tilde{A}}
\newcommand{\bphi}{\varphi}
\newcommand{\eps}{\varepsilon}
\newcommand{\calH}{\mathfrak H}
\newcommand{\calT}{\mathfrak T}
\newcommand{\calV}{\mathfrak V}
\newcommand{\calR}{\mathfrak R}
\newcommand{\E}{\mathbb E}
\newcommand{\N}{\mathbb N}
\newcommand{\Z}{\mathbb Z}
\newcommand{\R}{\mathbb R}
\newcommand{\cL}{\mathbb{L}}
\newcommand{\Var}{\textrm{Var}}
\newcommand{\qnot}{\mbox{\r q}}
\newcommand{\pnot}{\mbox{\r p}}
\newcommand{\tjo}{\overline{t}_j}
\newcommand{\tju}{\underline{t}_j}
\newcommand{\brk}[1]{\left(#1\right)}
\newcommand{\abs}[1]{\left|#1\right|}
\newcommand{\absdet}[1]{\left|\det #1\right|}
\newcommand{\pd}[2]{\frac{\partial#1}{\partial#2}}
\newcommand{\pdd}[2]{\frac{\partial^2#1}{\partial#2^2}}
\newcommand{\pds}[3]{\frac{\partial^2#1}{\partial#2\partial#3}}
\newcommand{\ud}[1]{\, \mathrm{d}#1}
\newcommand{\dsum}[3]{\sum_{\scriptstyle{#1} \atop \scriptstyle{#2}}^{#3}}

\makeatletter \@addtoreset{equation}{section} \makeatother

\newtheorem{Thm}{Theorem}[section]

\begin{document}


\title{Optimal Prediction in Molecular Dynamics}
\author{Benjamin Seibold \\
Department of Mathematics \\
Technical University of Kaiserslautern \\
Gottlieb-Daimler-Stra\ss e, 67653 Kaiserslautern, Germany \\
Email: {\tt seibold@mathematik.uni-kl.de}}
\date{}
\maketitle


\begin{abstract}
\noindent
Optimal prediction approximates the average solution of a large system
of ordinary differential equations by a smaller system. We present how
optimal prediction can be applied to a typical problem in the field of
molecular dynamics, in order to reduce the number of particles to be
tracked in the computations. We consider a model problem,
which describes a surface coating process, and show how asymptotic
methods can be employed to approximate the high dimensional conditional
expectations, which arise in optimal prediction. The thus derived
smaller system is compared to the original system in terms of statistical
quantities, such as diffusion constants. The comparison is carried out by
Monte-Carlo simulations, and it is shown under which conditions optimal
prediction yields a valid approximation to the original system.
\end{abstract}

\section{Introduction}

Computations in the field of molecular dynamics typically require a large
computational effort due to two factors:
\begin{enumerate}
\item
Small time steps are required to resolve the fast atomic oscillations.
\item
Large systems are obtained due to the large amount of atoms which have to
be computed.
\end{enumerate}
A wide variety of methods has been developed to remedy these problems.
Larger time steps are admitted e.g.~by smoothing algorithms, which
average in time over the fast oscillations. Various other methods
reduce the degrees of freedom, e.g.~multipole methods \cite{FMM} in the
context of long range particle interactions. In this paper we investigate
whether the method of optimal prediction, as presented and analyzed in
\cite{BCC,CHK1,CHK2,CHK3,CKK1,CKK2,CKK3,ChorinNotes,CKL,HK1,K1}, can
in principle be applied to problems in molecular dynamics in order to
reduce the number of atoms to be computed. As a first step in this
investigation we confine to a one dimensional model problem which
inherits particular properties from a real molecular dynamics problem.
In Section~\ref{sec:problem_description} we present the real problem as
it arises in applications, and derive the simplified model problem.
The considered problem is Hamiltonian, hence in
Section~\ref{sec:optimal_prediction} we present the method of optimal
prediction in the special case of Hamiltonian systems.
In Section~\ref{sec:op_applied} we apply the method of optimal prediction
to the model problem. This yields expressions involving high dimensional
integrals. We evaluate these integrals by asymptotic methods, employing
the fact that the process is running at a low temperature, which yields
a new and smaller system. Section~\ref{sec:speedup} deals with the numerical
speed-up. In Section~\ref{sec:comparing_systems} we define criteria, how
to investigate whether the new system is a valid approximation to the
original system. We show how important statistical quantities, such as
diffusion constants and energy fluctuations, can be obtained by numerical
experiments. These are presented in Section~\ref{sec:numerical_experiments},
and we investigate under which conditions optimal prediction preserves the
relevant statistical quantities. 

\section{Problem Description}
\label{sec:problem_description}

\subsection{The Physical Problem}

In the production of semiconductors a thin layer (a few atomic monolayers)
of copper has to be coated (sputtered) onto a silicon crystal. A technical
description of the process can be found in \cite{ProblemExperiment}. Important
for the quality of the product is that copper atoms must not penetrate too
deeply into the silicon crystal. The copper atoms penetrate firstly by their
kinetic energy when hitting the crystal surface, secondly by the process of
{\em atomic hopping}, which will be described in the following. In order to
obtain specific knowledge about these processes, molecular dynamics simulations
have to be carried out, as described in \cite{ProblemSimulationMD}.

One important aspect of the coating process is that the system is out of its
thermodynamical equilibrium only for very short times, namely for about
$10^{-11}$ seconds after one copper atom has hit the surface of the silicon
crystal. During this time the copper atom penetrates into the crystal and sonic
waves transport the impact energy away to the bottom of the crystal, which is
constantly being cooled. Hence, after $10^{-11}$ seconds the whole crystal is in
equilibrium again. On the other hand, the time between two copper atoms hitting
the crystal surface is on a time scale of $10^{-4}$ seconds, i.e.~the system is
in equilibrium nearly all the time, in particular the temperature is constant
with respect to space and time.

However, even in the state of thermodynamical equilibrium single copper atoms
can change their position in the silicon crystal by hopping events, i.e.~a copper
atom gains by accident enough energy to overcome the potential barrier between
two layers in the silicon crystal and thus hops to a neighboring cell. By atomic
hopping copper atoms can penetrate much deeper into the silicon crystal as their
impact energy would allow, hence the process in equilibrium cannot be omitted
from the computation. The average time between two hopping events is on a time
scale of $10^{-10}$ seconds, while the fast atomic oscillations happen on a time
scale of $10^{-14}$ seconds.

In this paper we show how the method of optimal prediction can be applied to
the system in equilibrium. Only the atoms at the top layers of the crystal,
where copper atoms can be found, are computed exactly, while the silicon atoms
in the lower layers are kept track of only in an average sense. In order to
keep the technical difficulties at a minimum, we set up a one dimensional model
problem which simulates atomic hopping.

\subsection{The Model Problem}
\label{subsec:model_problem}

In the model problem, we assume two major simplifications:
\begin{itemize}
\item
Focus on a one dimensional problem, i.e.~we consider $n$ atoms lined up like
beads on a cord. A single copper atom is inserted into a line of $n-1$ silicon
atoms.
\item
The potential $V(q)$ depends only on the pairwise distances of the
atoms, i.e.
\begin{equation}
V(q_1,\dots,q_n) = \dsum{i,j=1}{i<j}{n}f_\alpha(q_i-q_j).
\label{potential_energy}
\end{equation}
Here $\alpha\in\{1,2\}$, where $f_1$ is the potential between two silicon atoms
and $f_2$ is the potential between the copper atom and a silicon atom.
\end{itemize}

\begin{figure}[htb]
\begin{center}
\begin{minipage}[t]{.48\textwidth}
\includegraphics[width=.98\textwidth]{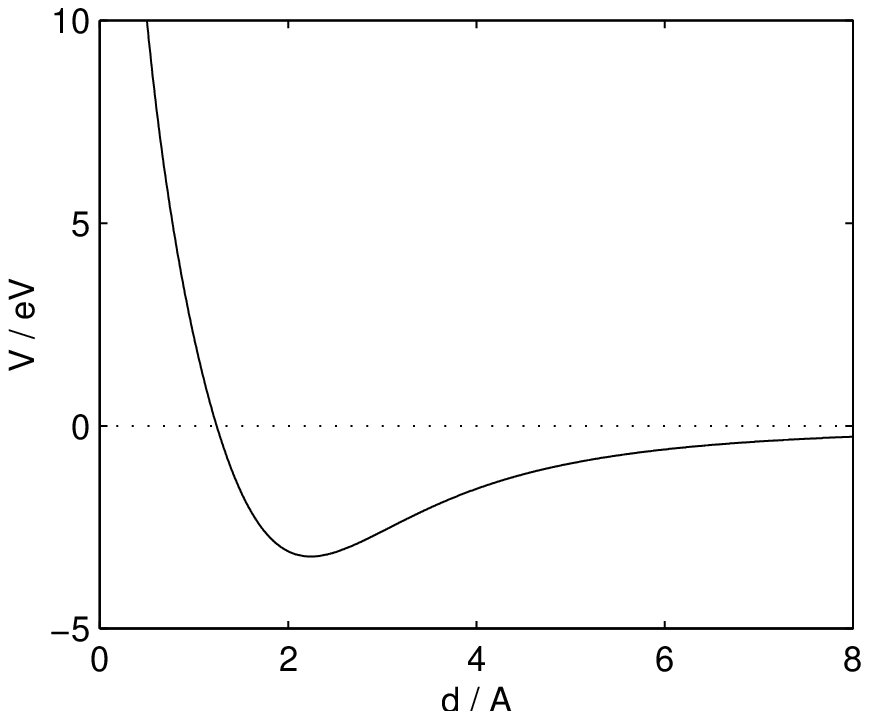}
\caption{Potential $f_1$ between two silicon atoms}
\label{plot_graph_pot_SiSi.eps}
\end{minipage}
\hfill
\begin{minipage}[t]{.48\textwidth}
\includegraphics[width=.98\textwidth]{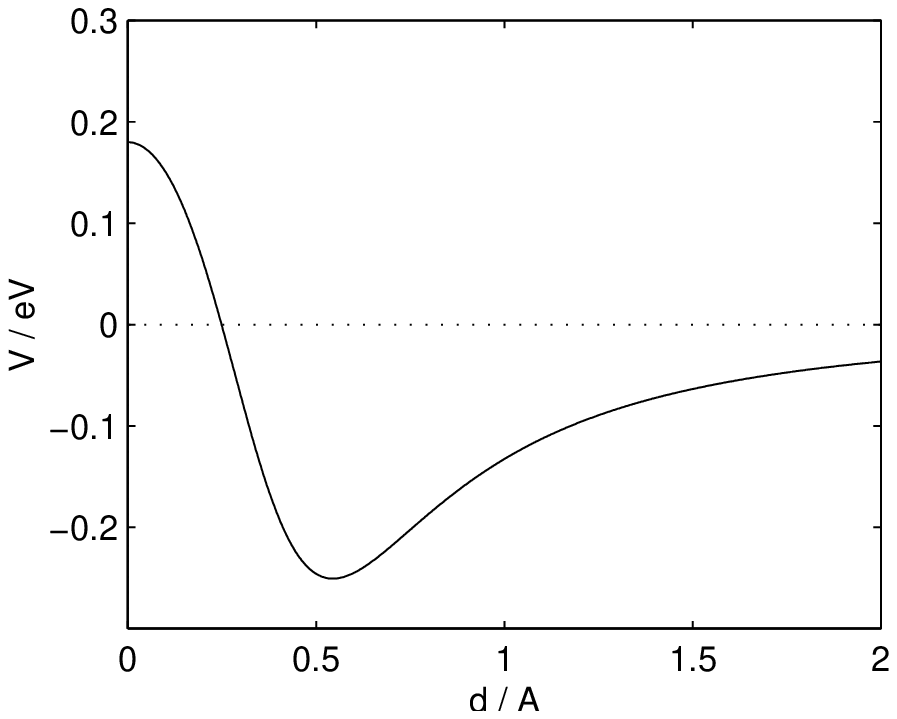}
\caption{Potential $f_2$ between copper and silicon}
\label{plot_graph_pot_CoSi.eps}
\end{minipage}
\end{center}
\end{figure}

Figures~\ref{plot_graph_pot_SiSi.eps} and \ref{plot_graph_pot_CoSi.eps} show the
pair potentials $f_1$ and $f_2$. The distance is given in {\AA}
($1\textrm{\AA}=10^{-10}$m) and the energy in eV
($1\textrm{eV}=1.6\cdot 10^{-19}$J).
The potentials are chosen to be close to the corresponding Lennard-Jones
potentials \cite{Hansen} in three space dimensions, in particular for $f_1$ the
position of the minimum ($r_e = 2.24\textrm{\AA}$) and the energy at the minimum
($E = -D$, where $D = 3.24\textrm{eV}$) are chosen to fit the correct values
given in \cite{Constants}. In reality, hopping between two silicon atoms happens
many times less likely than a copper hopping event. Hence, we neglect silicon
hopping in the model problem and choose $f_1$ to be infinite at the origin.

On the other hand, the pair potential $f_2$ between copper and silicon is
set to be finite at distance $0$, in order to allow copper hopping. While in
a three dimensional crystal a copper atom hops through one face of a crystal
cell, in one space dimension the copper atom can only hop directly over a
silicon atom in order to change its position in the crystal. Additionally,
the value at distance $0$ is significantly lowered compared to the potential
barrier set up by a face of a three dimensional silicon crystal. This makes
hopping events much more frequent and thus reduces the simulation time
required for observing hopping events.

In order to further increase the hopping rate, we increase the temperature
significantly. The real process is taking place at temperatures around
$500\textrm{K}$. We run the simulations at a temperature of $7000\textrm{K}$,
which is the maximum temperature that we can still call ``low''. In this
context a temperature being ``low'' means that the the dimensionless
quantity $\eps=\frac{k_B T}{D}$ is a small number. For real temperatures
one obtains $\eps \approx 0.01$, for the increased temperature $\eps = 0.13$.
Consequently, any result obtained for increased temperatures can be expected
to work even better at real temperatures.

In the following, we will always assume our system to behave as an ergodic
system. In particular this means that we assume space averages over a
number of atoms to be equal to time averages over a given time span.

\section{Optimal Prediction for Hamiltonian Systems}
\label{sec:optimal_prediction}

Optimal prediction was introduced by Chorin, Kast and Kupferman \cite{CKK1}
as a method to apply to {\em underresolved computation}, i.e.~to problems
which are computationally too expensive or where not enough data is at hand,
but prior statistical information is available. Sought is the mean solution
of a system, where only part of the initial data is known, and the rest is
sampled from an underlying measure. While the method is in principle not
restricted to a particular measure, in equilibrium statistical mechanics
one typically chooses the grand canonical distribution. Optimal prediction
approximates the mean solution by a new system which is smaller, and thus
cheaper to compute, than the original system. In this paper we will use a
simple optimal prediction system to consider only a smaller number of atoms
and ``averaging'' the other ones away. Note that optimal prediction is by
no means restricted to Hamiltonian systems, but for such systems it has
some particularly nice properties.

Consider a $2n$-dimensional Hamiltonian system of ordinary differential
equations
\begin{equation}
\dot q = \pd{H}{p},\quad \dot p = -\pd{H}{q}
\label{Hamiltonian_system}
\end{equation}
with the Hamiltonian function
\begin{equation}
H(q,p) = \frac{1}{2}p^2+V(q),
\end{equation}
representing an $n$-particle system in one space dimension. In the
following, we will consider the model problem in equilibrium, hence we assume
the position in state space to be distributed according to the grand
canonical distribution
\begin{equation}
f(q,p) = Z^{-1}e^{-\beta H(q,p)}.
\end{equation}
Here $\beta = \frac{1}{k_B T}$ is a constant with $k_B$ being the Boltzmann
constant, $T$ the (constant) temperature of the process, and
$Z= \int e^{-\beta H(p,q)}\ud{q}\ud{p}$ is a normalization constant.

We write the solution of (\ref{Hamiltonian_system}) as a phase flow, where
$\bphi(x,t)=\brk{q_1(t),p_1(t),\dots,q_n(t),p_n(t)}$ denotes the solution to
the initial condition $x=\brk{\qnot_1,\pnot_1,\dots,\qnot_n,\pnot_n}$.
Consequently, system (\ref{Hamiltonian_system}) can be rewritten as
\begin{eqnarray}
\frac{d}{dt}\bphi(x,t) &=& R(\bphi(x,t)), \label{equations_of_motion} \\ 
\bphi(x,0) &=& x. \nonumber
\end{eqnarray}
Assume that only $m$ of the $n$ atoms are of interest, which yields a
separation of the degrees of freedom into two groups $\bphi=(\hphi,\tphi)$,
where $\hphi = (\bphi_1,\dots,\bphi_{2m})$ represents the atoms of interest,
and $\tphi = (\bphi_{2m+1},\dots,\bphi_{2n})$ corresponds to the $n-m$ atoms
which should be considered only in an averaged sense. Let in the following
$l = n-m$ denote the number of averaged atoms. Typically, $m$ is significantly
smaller than $n$. Note that in our model problem silicon atoms cannot hop,
hence this separation stays valid over time, given the copper atom stays among
the silicon atoms of interest.

Now only part of the initial conditions $\hx = (x_1,\dots,x_{2m})$, namely the
ones corresponding to the variables which are of interest $\hphi$, are known,
while the other components $\tx = (x_{2m+1},\dots,x_{2n})$ are not known
exactly. Instead, for each choice of $\hx$ they are sampled from the conditioned
measure
\begin{equation}
f_{\hx}(\tx) = Z_{\hx}^{-1}e^{-\beta H(\hx,\tx)},
\label{conditioned_measure}
\end{equation}
where $Z_{\hx} = \int e^{-\beta H(\hx,\tx)}\ud{\tx}$ is the appropriate normalization
constant.
As in \cite{CHK2} we use the conditional expectation projection of a function $u(\hx,\tx)$
\begin{equation}
Pu = \E[u|\hx]
   = \frac{\int u(\hx,\tx)e^{-\beta H(\hx,\tx)}d\tx}{\int e^{-\beta H(\hx,\tx)}d\tx}.
\label{def_cond_exp}
\end{equation}
Optimal prediction as presented in \cite{BCC,CHK1,CHK2,CHK3,CKK1,CKK2,CKK3,CKL}
is interested in the first $2m$ components of the mean solution of
(\ref{equations_of_motion}), where the initial conditions $\hx$ are fixed and
$\tx$ are sampled from (\ref{conditioned_measure})
\begin{equation}
P\bphi(x,t) = \E[\bphi(x,t)|\hx]
= \frac{\int\bphi((\hx,\tx),t)e^{-\beta H(\hx,\tx)}d\tx}{\int e^{-\beta H(\hx,\tx)}d\tx}.
\label{mean_solution}
\end{equation}
For linear systems, expression (\ref{mean_solution}) can be computed exactly,
and it does not decay. In molecular dynamics, however, the Hamiltonian system
is in general nonlinear. As observed in \cite{BCC,CHK1}, for nonlinear systems,
the mean solution decays, which is interpreted as a loss of information as the
first $2m$ variables tend to the thermodynamical equilibrium. In \cite{BCC}
the authors give a deeper physical reasoning for the decay. An application of
the Mori-Zwanzig formalism as in \cite{CHK1,CHK2,CHK3} yields the formal
explanation.

For each choice of $\hx$ the mean solution (\ref{mean_solution}) can be
approximated by Monte-Carlo sampling, i.e.~sampling $N$ times $\tx$ from the
conditioned measure (\ref{conditioned_measure}), solving $N$ times the system
(\ref{equations_of_motion}) with initial values $(\hx,\tx)$, and averaging
over all solutions. Obviously, this is more expensive than solving the
original system itself.

In \cite{CHK1} the term {\em first order optimal prediction} has been assigned
to idea of applying the conditional expectation projection $P$ to the right
hand side $R$
\begin{equation}
\calR = PR = \E[R|\hx],
\label{PR}
\end{equation}
which yields a function of just $2m$ variables. Hence,
$\hat\calR = (\calR_1,\dots,\calR_{2m})$ is a function from $\R^{2m}$ to
$\R^{2m}$. The first order optimal prediction system is defined as
\begin{equation}
\dot y(t) = \hat\calR(y(t)), \quad y(0) = \hx.
\label{first_order_op}
\end{equation}
An important result, which allows to restrict to considering the Hamiltonian
function only, is the following
\begin{Thm}[O.~Hald \cite{ChorinNotes}]
\label{Thm_OP_Hamiltonian}
If a system is Hamiltonian, then its optimal prediction system is also Hamiltonian
with the Hamiltonian function
\begin{equation}
\calH(\hq,\hp) = -\frac{1}{\beta}\log\brk{\frac{1}{c}\int\int
e^{-\beta H(\hq,\hp,\tq,\tp)}\ud{\tq}\ud{\tp}}.
\label{op_hamiltonian}
\end{equation}
Here $c$ is a constant with unit $[c]=[\tq]\cdot[\tp]=\brk{kg\frac{m^2}{s}}^l$.
The exact value of $c$ is of no importance for the dynamics.
\end{Thm}

This theorem implies that for nonlinear problems first order optimal prediction
is never a good approximation for long times, since the mean solution decays,
i.e.~loses energy, while the first order optimal prediction system is Hamiltonian,
and thus energy preserving. In \cite{CHK1,CHK2,CHK3,ChorinNotes} {\em higher order
optimal prediction} methods have been derived, which reproduce the desired decay
of the mean solution.

However, in our case, we are not interested in the mean solution, but in a
$2m$-dimensional system, which yields the same behavior of the first $m$ atoms
as the full $2n$-dimensional system would have yielded. This does not
necessarily require a good approximation in the sense of trajectories
(which the mean solution focuses on), but the relevant statistical quantities
should be the recovered. In particular, the $2m$-dimensional system should be
Hamiltonian again. Hence, we choose the first order optimal prediction system
as the sought $2m$-dimensional system. Of course this choice can only be reasoned,
if it turns out that the relevant statistical quantities are indeed preserved.
We will focus on this question in
Sections~\ref{sec:comparing_systems}~and~\ref{sec:numerical_experiments}.

\section{Optimal Prediction Applied to the Model Problem}
\label{sec:op_applied}

\subsection{Appropriate Domains of Integration}
\label{subsec:domains_of_integration}

As the pair potentials $f_1$, $f_2$ vanish at infinity, the expression
\begin{equation}
\int_{\R^n}e^{-\beta H(x)}\ud{x}
\label{integral_over_canonical_measure}
\end{equation}
is not finite. Hence, the canonical measure $f(q,p) = Z^{-1}e^{-\beta H(q,p)}$
does not make sense as a probability distribution if the particle positions
are not restricted in some way. In many text books on statistical mechanics
the whole system is put into a box of finite volume,
i.e.~$\brk{q_1,\dots,q_n}\in [-L,L]^n$, where $L$ is suitably large. This
essentially means that all atoms
are trapped, but no ordering is specified. In our model problem, however, the
silicon atoms are ordered, and since silicon atoms cannot hop over each other,
this order stays valid over time. This is an information which we do not wish
to average out. Hence, we restrict the position vector $\brk{q_1,\dots,q_n}$
to the domain
\begin{equation}
M^L = \{(q_1,\dots,q_n) \in [-L,L]^n | q_2<\dots<q_n\}.
\label{domain_ML}
\end{equation}
Here $q_1$ is the position of the copper atom, which can be anywhere in
$[-L,L]$, as it can hop freely. The positions of the silicon atoms
$q_2,\dots,q_n$, however, are restricted to be ordered from left to right.
With respect to optimal prediction, the domain
\begin{equation}
M^L_{\hq} = \{(q_{m+1},\dots,q_n) \in [-L,L]^l | q_m<q_{m+1}<\dots<q_n\}
\end{equation}
has to be introduced. For each fixed position vector $\hq=(q_1,\dots,q_m)$
the other $l$ silicon atoms $\tq$ are restricted to be positioned right of
the first $m$ atoms, and ordered. This setup assumes that the copper always
stays among the first $m$ atoms.

\subsection{The Optimal Prediction Hamiltonian}
\label{subsec:op_hamiltonian}

When considering the optimal prediction Hamiltonian (\ref{op_hamiltonian})
as given by Theorem~\ref{Thm_OP_Hamiltonian}, one can easily check that
the kinetic and potential energy separate
\begin{equation}
\calH(\hq,\hp)
= \underbrace{-\frac{1}{\beta}\log\brk{\frac{1}{c_p^l}\int_{\R^l}
e^{-\beta T(\hp,\tp)}\ud{\tp}}}_{=\calT(\hp)}
\underbrace{-\frac{1}{\beta}\log\brk{\frac{1}{c_q^l}\int_{M^L_{\hq}}
e^{-\beta V(\hq,\tq)}\ud{\tq}}}_{=\calV(\hq)}.
\label{op_hamiltonian_splitted}
\end{equation}
Here $c_p$ and $c_q$ are constants with units
$[c_p] = \textrm{kg}\frac{\textrm{m}}{\textrm{s}}$ and $[c_q] = \textrm{m}$.
Since $T=\frac{1}{2}\sum_{i=1}^{n}\frac{p_i^2}{m_i}$, the first term of
(\ref{op_hamiltonian_splitted}) can be computed directly as
\begin{equation}
\calT(\hp) = \frac{1}{2}\sum_{i=1}^{m}\frac{p_i^2}{m_i}+C,
\end{equation}
where the constant
$C=-\frac{1}{2\beta}\sum_{i=m+1}^{n}\log\brk{\frac{2\pi m_i}{\beta c_p^2}}$
is of no relevance for the dynamics. Hence, with respect to the momenta
applying first order optimal prediction means omitting the momenta
$p_{m+1},\dots,p_n$. For the potential $V$, however, life is far from being
as easy as for the kinetic energy $T$, as the $q$-variables do not separate
and are no quadratic functions. Thus, an analytic evaluation of the first
order optimal prediction potential $\calV(\hq)$ is in general impossible,
or at least too complicated to be of any use. Hence, we employ an asymptotic
expansion of $\calV(\hq)$.

\subsection{Low Temperature Asymptotics}

The dimensionless quantity $\eps = \frac{1}{D\beta} = \frac{k_B T}{D}$ is
small for low temperatures. The optimal prediction potential expressed in
terms of the quantity $\eps$ is
\begin{equation}
\calV(\hq) = -D\eps\log\brk{\frac{1}{c_q^l}\int_{M^L_{\hq}}
e^{-\frac{1}{\eps} \tV(\hq,\tq)}\ud{\tq}},
\label{OP_V_normalized}
\end{equation}
where $\tV(\hq,\tq) = \frac{1}{D}V(\hq,\tq)$ is the potential normalized in
such a way, that the potential of two atoms at equilibrium distance has the
value -1.

Using {\em Laplace's method for integrals of real variables}
\cite{Murray,Olver}, one can find an asymptotic approximation to
(\ref{OP_V_normalized}) for $\eps$ small. In some textbooks this method is
also referred to as {\em Watson Lemma}. Assume for the moment that for a
fixed choice of $\hq$ the function $\tV(\hq,\tq)$ has a unique global
minimizer $r(\hq)\in\R^l$ with respect to $\tq$, and that the Hessian at
this point $\pdd{\tV}{\tq}(\hq,r(\hq))$ is regular. In the following
derivation we use the abbreviatory notations $r=r(\hq)$ and
$H_{\tq}\tV=\pdd{\tV}{\tq}(\hq,r(\hq))$. Laplace's method approximates
$\tV(\hq,\tq)$ by a quadratic function located at the minimum, yielding
the following asymptotic approximation for $\eps\to 0$
\begin{eqnarray}
\int_{M^L_{\hq}} e^{-\frac{1}{\eps}\tV(\hq,\tq)}\ud{\tq}
&\!\approx& \! \int_{M^L_{\hq}} e^{-\frac{1}{\eps}\brk{\tV(\hq,r)+\frac{1}{2}(\tq-r)^T
\cdot H_{\tq}\tV\cdot(\tq-r)}}\ud{\tq} \nonumber \\
&\!\approx& \! e^{-\frac{1}{\eps}\tV(\hq,r)}\cdot\int_{\R^l} e^{-\frac{1}{2\eps}\brk{(\tq-r)^T
\cdot H_{\tq}\tV\cdot(\tq-r)}}\ud{\tq}. \nonumber
\end{eqnarray}
Extending the set of integration to the whole $\R^l$ is valid, since the
minimum is always in the interior of $M^L_{\hq}$, provided $L$ is large
enough (see \cite{Olver}). Since $H_{\tq}\tV$ is assumed to be regular,
the transformation rule yields
\begin{equation}
\frac{1}{c_q^l}
\int_{\R^l}e^{-\frac{1}{2\eps}(\tq-r)^T\cdot H_{\tq}\tV
\cdot(\tq-r)}\ud{\tq}
= \sqrt{\frac{(2\pi\eps)^l}{\absdet{c_q^2 H_{\tq}\tV}}}.
\end{equation}
Given $\tV$ is of class $C^2$ (which is the case in our model problem), the
complete asymptotic expansion including the error term is
\begin{eqnarray}
\frac{1}{c_q^l}\int_{M^L_{\hq}}\!\! e^{-\frac{1}{\eps}\tV(\hq,\tq)}\ud{\tq}
= \sqrt{\frac{(2\pi\eps)^l}{\absdet{c_q^2 H_{\tq}\tV}}}
\!\cdot\! e^{-\frac{1}{\eps}\tV(\hq,r)} 
+e^{-\frac{1}{\eps}\tV(\hq,r)}\!\cdot\! O\brk{\eps^{\frac{l}{2}+1}},
\label{asymptotic_expansion_inside}
\end{eqnarray}
which follows directly from the one dimensional case as shown
in \cite[pp.~33-34]{Murray}. Substituting
(\ref{asymptotic_expansion_inside}) into (\ref{OP_V_normalized}), and
employing that $\log(1+x)\sim x$ as $x\to 0$ yields
\begin{equation}
\calV(\hq)
= V(\hq,r)+C+\frac{D\eps}{2}\log\absdet{c_q^2 H_{\tq}\tV}+O\brk{\eps^2},
\end{equation}
where the constant $C = -\frac{1}{2}Dl\eps\log\brk{2\pi\eps}$ is of no
relevance for the dynamics. Hence, we found a zeroth and a first order
asymptotic expansion in $\eps$ for $\calV$, which -- returning to long
notation -- are
\begin{eqnarray}
\calV_0(\hq) &=& V(\hq,r(\hq)),
\label{asymptotic_expansion_zeroth_order} \\
\calV_1(\hq) &=& V(\hq,r(\hq))+\eps\cdot
\frac{D}{2}\log\absdet{\frac{c_q^2}{D}\pdd{V}{\tq}(\hq,r(\hq))}.
\label{asymptotic_expansion_first_order}
\end{eqnarray}
Note that $\calV_0$ and $\calV_1$ approximate $\calV$ only up to constants,
which are irrelevant for the acting forces. Whenever in the following we speak
of $\calV_i$ approximating $\calV$, we always mean: ``up to constants''.

The above assumptions, that $\tV(\hq,\tq)$ has a unique global minimizer
$r(\hq)$ with respect to $\tq$, and that the Hessian at this point
$\pdd{\tV}{\tq}(\hq,r(\hq))$ is regular, can be observed to be guaranteed for
our model problem, given one restricts to the domain $M^L_{\hq}$. Both
assumptions can be relaxed for the zeroth order expansion, i.e.~also in the
case of multiple minimizers or a singular Hessian the zeroth order expansion
stays valid.

\subsection{Zero Temperature Limit}
\label{subsec:zero_temperature_limit}

The zeroth order approximation $\calV_0$
(\ref{asymptotic_expansion_zeroth_order}) is the limit of $\calV$
(\ref{OP_V_normalized}) as $\eps\to 0$, i.e.~$T\to 0$. Hence, we call
$\calV_0$ the \emph{zero temperature limit potential}. Since the dynamics
takes place at low temperatures, one can expect the correct optimal
prediction potential function $\calV$ to be close to the zero temperature
limit potential $\calV_0$. Hence, we run the low temperature optimal
prediction dynamics, which would be correctly described by $\calV$, with
the zero temperature limit dynamics given by $\calV_0$. We do not consider
the first order approximation (\ref{asymptotic_expansion_first_order}) here,
since the Hessian $\pdd{\tV}{\tq}(\hq,r(\hq))$ cannot be included in a
straightforward manner into the following derivation. The results in
Subsection~\ref{subsec:energy_fluctuations}, however, will show that further
investigation on the first order expansion could be worthwhile.

By going over from $\calV$ to $\calV_0$ we have formally replaced an
$l$-dimensional integration by an $l$-dimensional minimization problem.
At first glance this is no real improvement, since high dimensional global
minimization is also computationally expensive (see \cite{Annealing}).
However, the $l$-dimensional minimization means nothing else than placing
$l$ further atoms, such that the total potential energy is minimized.
Since $\calV_0$ is formally $m$-dimensional, we call the $l$ new atoms 
\emph{virtual atoms}. The restriction to the domain $M^L_{\hq}$ in
Subsection~\ref{subsec:domains_of_integration} guarantees that the $l$
virtual atoms are separated from the $m$ ``real'' atoms.

\subsection{Equations of Motion}
\label{subsec:equations_of_motion}

The zero temperature limit Hamiltonian is
\begin{equation}
\calH_0(\hq,\hp) = \calT_0(\hp)+\calV_0(\hq)
= \frac{1}{2}\hp^T\mathfrak{M}^{-1}\hp+V(\hq,r(\hq)),
\end{equation}
where $\mathfrak{M}$ is a diagonal matrix containing the masses $m_i$ of
the atoms $(\mathfrak{M})_{ii} = m_i$. In the following we assume $V(\hq,r)$
and $r(\hq)$ to be of class $C^1$. This allows to compute
\begin{eqnarray}
\pd{\calH_0}{\hp}(\hq,\hp) &\!\!=& \!\! \pd{\calT_0}{\hp}(\hp)
= \mathfrak{M}^{-1}\cdot\hat p,
\label{equations_of_motion_p} \\
\pd{\calH_0}{\hq}(\hq,\hp) &\!\!=& \!\! \pd{\calV_0}{\hq}(\hq)
= \pd{V}{\hq}(\hq,r(\hq))+\underbrace{\pd{V}{r}(\hq,r(\hq))}_{=0}
\cdot\frac{dr}{d\hq}(\hq)
= \pd{V}{\hq}(\hq,r(\hq)).
\label{equations_of_motion_q}
\end{eqnarray}
Note that $\pd{V}{r}(\hq,r(\hq))$ is zero, since $r(\hq)$ is the minimizer
of $V(\hq,r(\hq))$. Still, expression (\ref{equations_of_motion_q}) involves a
minimization problem, in order to place the virtual atoms. We circumvent the
minimization by deriving equations of motion for the virtual atoms, too. In
order to obtain the expression $\pd{r}{\hq}(\hq)$, we define
\begin{equation}
v(\hq) := \pd{V}{\tq}(\hq,r(\hq)).
\end{equation}
Since $r(\hq)$ is always chosen to minimize $V$, we have that
$v(\hq)=0~\forall\hq$. Thus
\begin{equation}
0 = \pd{v}{\hq}(\hq)
= \pds{V}{\hq}{\tq}(\hq,r(\hq))+\pdd{V}{\tq}(\hq,r(\hq))\cdot\pd{r}{\hq}(\hq).
\label{minimize_V}
\end{equation}
Solving (\ref{minimize_V}) for $\pd{r}{\hq}(\hq)$ yields an expression 
which can be substituted into the time evolution
$\frac{d}{dt}r(\hq) = \pd{r}{\hq}(\hq)\cdot\frac{d\hq}{dt}$.
This yields a closed system for the zero temperature limit optimal
prediction dynamics
\begin{eqnarray}
\frac{d}{dt}\hq &=& \mathfrak{M}^{-1}\cdot\hat p \nonumber \\
\frac{d}{dt}\hp &=& -\pd{V}{\hq}(\hq,r(\hq)) \label{complete_equations_of_motion} \\
\frac{d}{dt}r &=& -\brk{\pdd{V}{\tq}(\hq,r(\hq))}^{-1}\cdot\pds{V}{\hq}{\tq}(\hq,r(\hq))
\cdot\mathfrak{M}^{-1}\cdot\hat p, \nonumber
\end{eqnarray}
where initially $\hq_i(0) = q_i(0)$ and $\hp_i(0) = p_i(0)$ and $r(0)$ is
chosen such that $V(\hq(0),r(0))$ is minimal, which can be resolved by a
few Newton steps.

Since the virtual atoms always follow the minimum of the potential energy,
no momenta are required to describe their movement. Hence, system
(\ref{complete_equations_of_motion}) is just $(n+m)$-dimensional, instead of
$2n$-dimensional. It is a closed system of ordinary differential equations, and
the right hand side requires no integration or minimization. Still, there is an
$l$-dimensional linear system of equations to be solved. At this point, we can
employ the special structure of the potential energy (\ref{potential_energy})
in our problem.

Assume that the pair potentials $f_1$, $f_2$ reach over only $k$ (in our model
problem $k\approx 10$) equilibrium distances $d_0$. In the following derivation,
we assume the potentials to really vanish at greater distances. The derived
results hold approximately also for potentials which are negligibly small at
greater distances. Consequently, only the first $k$ virtual atoms
$r_{m+1},\dots,r_{m+k}$ actually have to be computed. The others will align
equidistantly right to the first $k$ ones. Since one atom has no influence on
atoms more than $k$ equilibrium distances away, only $l=2k$ virtual atoms need
to be considered, where the last $k$ ones are aligned equidistantly. As we are
interested in the case $m\ll n$, enough virtual atoms are present, such that
boundary effects can be neglected. Let the positions of the $2k$ virtual atoms
be denoted by
\begin{equation}
r = \brk{\begin{array}{c} r^V \\ r^E \end{array}},
\end{equation}
where $r^V = \brk{r_{m+1},\dots,r_{m+k}}^T$ and
$r^E = r_{m+k}e+\brk{d_0,2d_0,\dots,kd_0}^T$ with $e=\brk{1,\dots,1}^T$.
The time derivative is
\begin{equation}
\frac{d}{dt}r = \brk{\begin{array}{c} \dot r^V \\ \dot r_{m+k}e \end{array}}.
\end{equation}
In this setup the matrices in (\ref{complete_equations_of_motion}) take a
special structure:
\begin{itemize}
\item
The Hessian $\pdd{V}{\tq}(\hq,r(\hq))$ is a diagonal band matrix with band
width $k$. In block form, where each block is of size $k\times k$, it is
\begin{equation}
\pdd{V}{\tq}(\hq,r(\hq)) =
\brk{\begin{array}{cc}
A_{11} & A_{12} \\
A_{21} & A_{22}
\end{array}},
\end{equation}
where $A_{12}$ is lower triangular and $A_{21}$ upper triangular.
\item
The matrix $\pds{V}{\hq}{\tq}(\hq,r(\hq))$ is upper triangular with width
$k$, i.e.
\begin{equation}
\pds{V}{q_i}{q_{m+j}}(\hq,r) 
= f''(r_j-q_i)
\begin{cases}
= 0 & \text{if } |m+j-i| > k \\
\neq 0 & \textrm{else}
\end{cases}
\end{equation}
In block form it can be written as
\begin{equation}
\pds{V}{\hq}{\tq}(\hq,r(\hq)) =
\brk{\begin{array}{cc}
0 & B_{12} \\
0 & 0
\end{array}},
\end{equation}
where $B_{12}$ is an upper triangular $k\times k$ matrix.
\end{itemize}
Substituting these special vectors and matrices into the equation for the
virtual atoms in (\ref{complete_equations_of_motion}) yields the following
equation of motion in block form
\begin{equation}
\brk{\begin{array}{cc}
A_{11} & A_{12} \\
A_{21} & A_{22}
\end{array}}
\cdot
\brk{\begin{array}{c} \dot r^V \\ \dot r_{m+k}e \end{array}}
=
\brk{\begin{array}{cc}
0 & B_{12} \\
0 & 0
\end{array}}
\cdot
\mathfrak{M}^{-1}\cdot\hat p,
\end{equation}
which implies the $k$-dimensional system for $r^V$
\begin{equation}
\bar A_{11}\cdot\dot r^V = B_{12}\cdot\mathfrak{M}_l^{-1}\cdot\hat p_l.
\end{equation}
Here $\bar A_{11} = A_{11}+\brk{A_{12}\cdot e}\cdot e_k^T$, where
$e_k = \brk{0,\dots,0,1}^T$. The momenta of the last $k$ real atoms are
denoted by $\hat p_l=\brk{p_{m-k+1},\dots,p_m}^T$, and $\mathfrak{M}_l$ is the
diagonal matrix containing the corresponding masses.
This relation can be interpreted as a boundary layer condition which acts as if
the crystal of silicon atoms was continued to infinity, although it is actually
cut off after the $m$-th atom.

With the above modifications the zero temperature limit optimal prediction
system (\ref{complete_equations_of_motion}) becomes an explicit
$(2m+k)$-dimensional system of equations. Hence, one can expect this system
to yield a computational speed-up, depending on the values of $n$, $m$ and $k$.
The question of speed-up will be considered in Section~\ref{sec:speedup}.
Due to the various approximations in achieving the smaller system it is not
at all clear whether the smaller system yields the same dynamics as the original
system. In Section~\ref{sec:comparing_systems} we will compare the new system
to the original system and investigate under which conditions the new system
reflects the ``correct'' dynamics, and under which conditions it does not.

\section{Computational Speed-Up}
\label{sec:speedup}

Besides the above physical interpretations, the actual intention was to
use optimal prediction as a method to reduce the computational effort.
In this section, we consider both the version with $l$ virtual atoms and
the boundary layer version.
We compare the CPU times for computations of the two optimal prediction
systems with the CPU time for the corresponding computations of the
original system. The comparison is carried out in dependence of the sizes
$n$ and $m$. Since the original version of optimal prediction does only
replace real atoms by virtual ones, while on the other hand the boundary
layer condition version allows to really omit atoms, a significant speed-up
can only be expected from the boundary layer condition version.

\begin{figure}[htb]
\begin{center}
\begin{minipage}[b]{.48\textwidth}
\includegraphics[width=.98\textwidth]{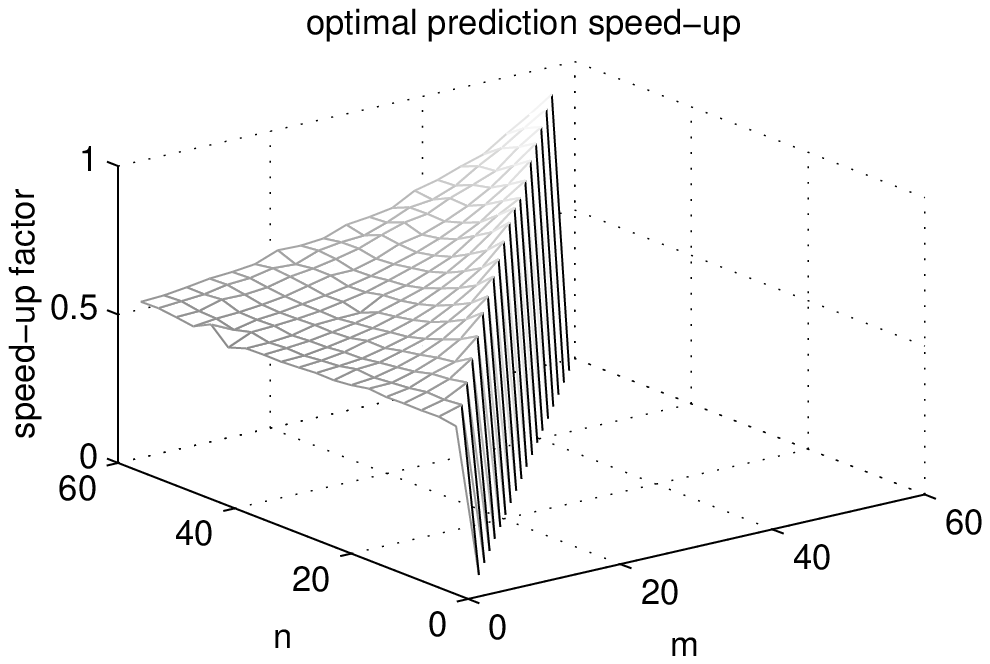}
\includegraphics[width=.98\textwidth]{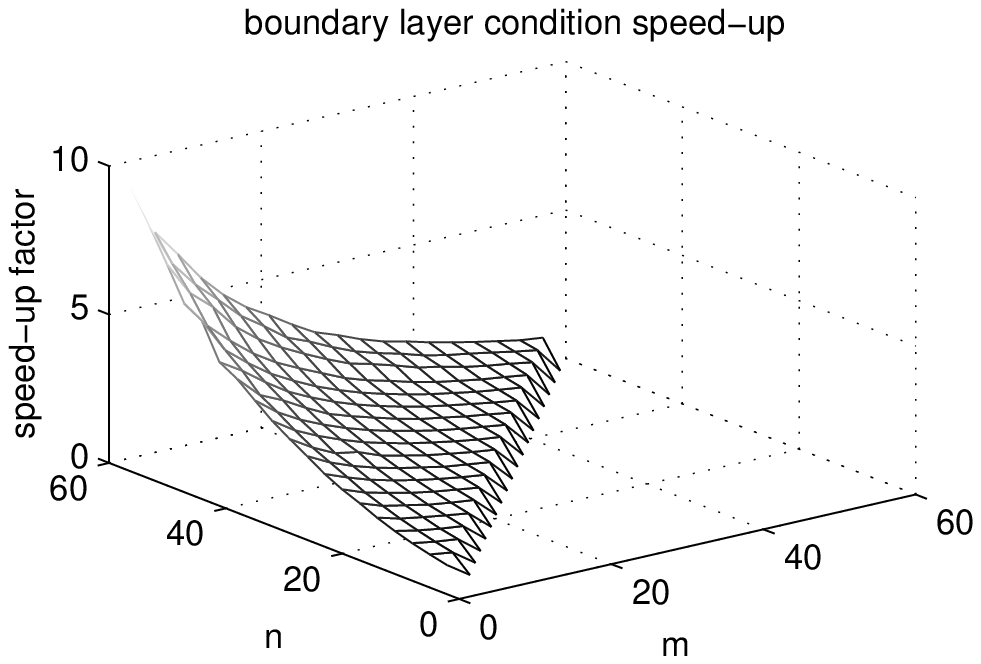}
\caption{Speed-up factors depending on $n$ and $m$.}
\label{speedup_factors.eps}
\end{minipage}
\hfill
\begin{minipage}[b]{.48\textwidth}
\includegraphics[width=.98\textwidth]{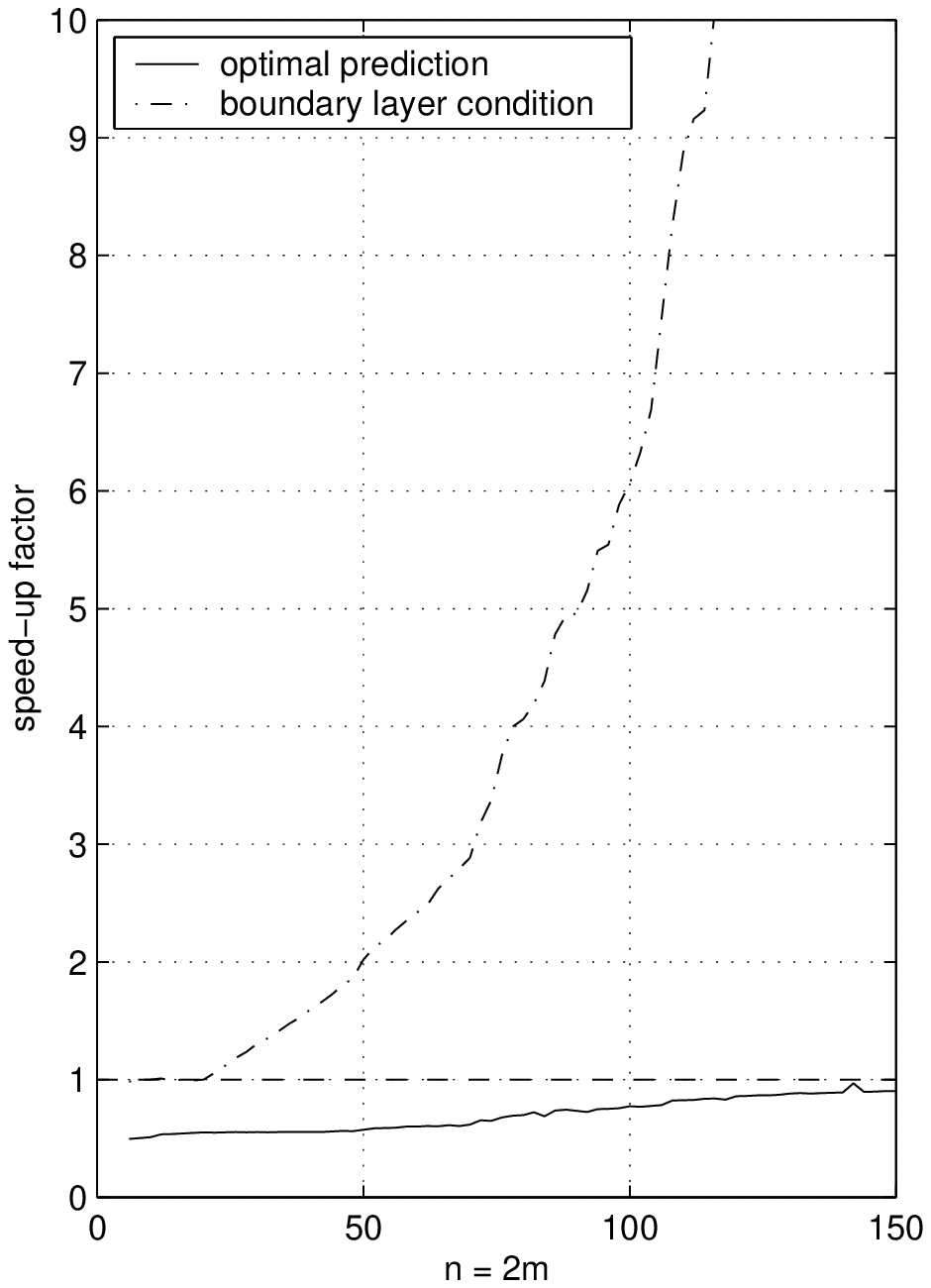}
\caption{Speed-up factors for $n=2m$}
\label{speedup_factors_half.eps}
\end{minipage}
\end{center}
\end{figure}

The computations were performed on a network of AMD Athlon-6 1.4 GHz
computers. The boundary layer condition version was computed with $k=10$
virtual atoms. Figure~\ref{speedup_factors.eps} shows the speed-up factors
with respect to the original system for the two versions of optimal
prediction. The original version of optimal prediction does not yield any
acceleration (the speed-up factors are less than 1). Apparently, for our
model problem setting up and solving the linear system in
(\ref{complete_equations_of_motion}) is more expensive than computing the
full system of equations. The boundary layer condition version, on the other
hand, yields significant speed-up factors for small $m$. In principle, one
can achieve arbitrarily high speed-up factors by keeping $m$ and $k$ fixed
and increasing $n$. In most cases, however, the original crystal size $n$
is given a priori, and suitable values for $m$ are given by the
requirement that the new system must have the same dynamics as the original
one, as discussed in Section~\ref{sec:comparing_systems}.

A choice of $m$ which is reasonable in many cases, is cutting the whole
crystal of $n$ atoms into two halves, i.e.~$m=\lfloor\frac{n}{2}\rfloor$. The
speed-up factors for the two optimal prediction systems are shown in
Figure~\ref{speedup_factors_half.eps}. While the original version does not
reduce the computational effort, the boundary layer condition version
yields a good speed-up for larger $n$.

Care should be taken when generalizing these results. In our model problem
the potential functions are fairly cheap to evaluate. In more complicated
systems it could very well pay off to solve the linear system in
(\ref{complete_equations_of_motion}) instead, or to solve the minimization
problem directly. On the other hand, the boundary layer condition version
of optimal prediction could possibly fail to work in other applications,
e.g.~in three space dimensions. Such questions will have to be investigated
when applying optimal prediction to a new problem.

\section{Comparing Two Molecular Dynamics Systems}
\label{sec:comparing_systems}

In molecular dynamics trajectories in high dimensional phase space are no
appropriate means for comparing two systems, since initial positions and
momenta can never be known exactly, and molecular dynamics is typically
chaotic. Instead, ``comparing'' means to test whether both systems have
similar dynamics. This is represented by statistical quantities such as
time correlation functions, diffusion constants, fluctuations of energy,
etc. We consider the following statistical processes in order to compare
the two systems:
\begin{itemize}
\item {\bf The distribution of the position the copper atom.}
A copper atom, which is initially located always at the same position, is
in the ensemble of many experiments described by a diffusion process, whose
distribution can be approximated by Monte-Carlo sampling.
\item {\bf The number of hopping events up to a fixed time.}
\item {\bf The fluctuation of energy of the first $\mathbf{m}$ atoms.}
The energy of the first $m$ atoms fluctuates around some fixed average.
We consider the variance of the energy over a fixed time interval.
\end{itemize}
The first two quantities are related to the diffusion of a single copper
atom in the silicon crystal due to hopping events. Since copper hopping has
been the effect of interest in the first place, it is a natural criterion
for comparison. 

In molecular dynamics, statistical quantities of ergodic systems can be
computed by long time averaging or Monte-Carlo sampling. ``Long time
averaging'' means running a single computation and using limiting
processes to approximate statistical quantities. Important examples,
e.g.~for approximating self-diffusion constants \cite{CageCorr}, are
the {\em Green-Kubo formula} \cite{Hansen,FicksLaw} or the {\em Einstein
relation}, which both approximate ensemble averages for ergodic systems
by long time averaging. In our application, however,
long-time computations are problematic, since firstly the copper atom may
travel to the boundaries of the silicon crystal and secondly sonic waves
will come in effect, as shown in Section~\ref{sec:numerical_experiments}.
Hence, we obtain the diffusion constant by Monte-Carlo sampling instead,
i.e.~we solve the same system over and over again with short computation
times, where the initial conditions are sampled from the canonical
measure. In other words: We obtain the averaged quantities not by
long-time averaging, but by averaging over many samples.

Sampling both the initial positions $q_i$ and momenta $p_i$ from the
canonical measure $Z^{-1}e^{-H(q,p)}$ would require expensive methods
like {\em acceptance-rejection methods} or {\em Metropolis sampling}
\cite{Devroye,Hammersley} due to the structure of the potential $V(q)$.
We circumvent such problems by setting the initial positions $q_i$
into the potential minimum and sampling the initial momenta $p_i$ from
$\tilde Z^{-1}e^{-T(p)}$, which is just sampling independently from Gaussian
distributions. In our simulations after about $5\cdot 10^{-14}\textrm{s}$
the Hamiltonian dynamics has driven the system into equilibrium. Additionally,
keeping the initial positions fixed automatically guarantees to remain in the
correct domain $M^L$ as given by (\ref{domain_ML}).

\subsection{A Random Walk Model for the Copper Diffusion}
\label{subsec:model_copper_diffusion}

The considered copper diffusion is due to hopping events, while displacements
due to short oscillations between the same two silicon atoms are not taken
into account. Hence, the corresponding diffusion process is discrete in space
on the spatial grid $\{-\nu d_0,\dots,0,\dots,\nu d_0\}$. Let us assume for
the moment, that the diffusion process is linear and hence described by a
$(2\nu+1)$-dimensional {\em compartment model}
\begin{equation}
\dot u(t) = A\cdot u(t), \quad u(0) = e_{m+1} = \brk{0,\dots,0,1,0,\dots,0}^T,
\label{CM} 
\end{equation}
where $A\in \R^{\brk{2\nu+1}\times\brk{2\nu+1}}$ has column sums equal to
zero to ensure mass conservation, i.e.~$\frac{d}{dt}\sum_i u_i(t) = 0$. The
analytical solution to (\ref{CM}) is
\begin{equation}
u(t) = \exp(tA)\cdot e_{m+1}.
\label{CM_solution}
\end{equation}
One example for such a process is given by the tridiagonal Toeplitz (up to
the boundary entries) matrix with the stencil
$\frac{\kappa}{d_0^2}\brk{1,-2,1}$, which is a finite difference
approximation to the heat equation. Hence, the compartment model (\ref{CM})
with this matrix converges to the heat equation as $d_0\to 0$. Still, there
are many other matrices $A$, whose discrete diffusion processes all converge
to the heat equation as $d_0\to 0$.

The hopping behavior of the copper atom can be modeled as a specific random
walk. One important aspect of the copper hopping is that hopping events are
{\em correlated}, in the sense that given the copper atom has just hopped,
it is much more likely than normally that a second hopping event to the same
direction follows, since the kinetic energy is not completely lost in one
single jump. Hence, the copper diffusion is formally non-Markovian.
Still, the hopping can be described as a Markovian random walk by assuming
hopping events to be uncorrelated, but to allow the copper atom to hop over
more than one silicon atom in one single hopping event.

These assumptions lead to a model which is typical in the context of
stochastic processes, see~\cite{Lawler}. Let $X_t$ denote the position
of the copper atom at time $t$. For the $n^\mathrm{th}$ hopping event let
$T_n$ be the hopping time, $\Delta T_n = T_n-T_{n-1}$ the time since the
previous hopping event, and $\Delta_n \in \Z\setminus \{0\}$ the number
of silicon atoms which the copper atoms hops over to the right. Here
$\Delta_n<0$ means hopping to the left. Assume now that both the
$\Delta T_n$ and the $\Delta_n$ are independent and distributed according
to
\begin{eqnarray}
&&P\brk{\Delta T_n \in [s,s+\ud{s})} = \alpha\cdot\exp(-\alpha s)\ud{s},
\label{time_between_hopping_events} \\
&&P\brk{\abs{\Delta_n} = i} = p_i.
\end{eqnarray}
Here $\alpha$ is a parameter controlling the hopping rate (see
\cite{Krengel,Lawler} for a derivation), and $(p_i)_{i\in\N}$ is a
non-negative sequence satisfying $\sum_{i\in\N}p_i = \frac{1}{2}$ and the
constant $\gamma = \sum_{i=1}^k (d_0i)^2 p_i$ is finite. For our model
problem we simply assume $p=\brk{p_1,\dots,p_k}$. Additionally, we restrict
to symmetric random walks, which is reasonable as long as the copper atom
does not approach the crystal boundaries.

An analysis of the described random walk, following the analysis in
\cite{Lawler}, yields that the variance of the process at time $t$ equals
$\Var(X_t) = 2\gamma\alpha t$. A comparison with the variance given by the
heat equation yields the relation
\begin{equation}
\kappa = \gamma\alpha,
\label{relation_kappa_alpha}
\end{equation}
which allows us to speak consistently of a diffusion constant $\kappa$ also
for the discrete diffusion process (\ref{CM}). As derived in \cite{Lawler}
the probability distribution of the random walk $X_t$ is described by
(\ref{CM}), where the so called {\em infinitesimal generator} $A$ is a band
matrix with the stencil $\alpha\cdot\brk{p_k,\hdots,p_1,-1,p_1,\hdots,p_k}$.
Define $\tilde A$ as a corresponding band matrix with the stencil
$\gamma^{-1}\cdot\brk{p_k,\hdots,p_1,-1,p_1,\hdots,p_k}$, hence $A = \kappa\tA$.
Both matrices have to be changed in the upper and lower rows according to
the appropriate boundary conditions.

Assume now, that the values $p_1,\dots,p_k$ are known. Hence, the matrix
$\tilde A$ is completely determined. Let $v(t)\in\R^{2\nu+1}$ denote the
numerically obtained distribution vectors for the position of the copper
atom for all times $t$. Initially $v(0) = e_{m+1}$, as in (\ref{CM}). Since
it is unclear, whether the diffusion parameter $\kappa$ is constant in time,
we let it be a time dependent parameter $\kappa(t)$, which we compute at
times $t_1,\dots,t_\mu$. For obtaining the value $\kappa_j = \kappa(t_j)$
we consider the diffusion on the time interval $I_j = [\tju,\tjo]$, where
$\tju = t_j-\frac{\Delta t}{2}$ and $\tjo = t_j+\frac{\Delta t}{2}$. Given
$\Delta t$ is not too large, we can approximately assume the diffusion
parameter to be constant on the given interval:
$\kappa(\tau)=\kappa_j \forall \tau\in I_j$. On $I_j$ we use the data
provided by the real evolution $v(t)$ to compute an $\cL^2$-fit of the
diffusion parameter $\kappa_j$ with respect to error functional
\begin{equation}
F(\kappa_j) = \int_{\tju}^{\tjo}
              \|e^{(\tau-\tju)\kappa(t_j)\tA}
	      \cdot v(\tju)-v(\tau)\|_2^2\ud{\tau},
\label{L2_error}
\end{equation}
which is particularly stable with respect to errors in $v$ due to the
Monte-Carlo sampling. The time $\Delta t$ must on the one hand be small enough
to justify the approximation that the diffusion parameter is constant on the
intervals $I_j$, on the other hand it must be large enough to have already some
diffusion taken place. In Subsection~\ref{subsec:diffusion_parameters} we
compute diffusion parameters for the numerical data obtained for our model
problem.

\section{Numerical Experiments}
\label{sec:numerical_experiments}

We simulate a crystal of $n=70$ atoms with $69$ silicon atoms and $1$ copper
atom, which starts at the $22^\mathrm{nd}$ position. For the optimal
prediction system, we choose $m=50$. These are enough atoms such that
averaged quantities make sense and one can speak of a thermodynamical
equilibrium. Additionally, the crystal has a reasonable interior region
which is not affected by any boundary effects. The integration is performed
by the classical explicit fourth order Runge-Kutta method, which we prefer
over energy preserving methods in this context, as it allows significantly
larger time steps, while still resolving the hopping events accurately.
The integration time step is $\Delta t=2.5\cdot 10^{-15}\textrm{s}$, and
the computation time is $t^*=4.0\cdot 10^{-13}\textrm{s}$, which is short
enough, such that the change in total energy due to the integrator is
insignificant. We solve the system $N=25000$ times at a temperature
$T=7000\textrm{K}$, with the initial data sampled as described in
Section~\ref{sec:comparing_systems}. For our experiments, Monte-Carlo
estimates yield an expected error in the distribution of about
$4\cdot 10^{-3}$, which is significantly smaller than the difference of
any two quantities in comparison.

Note that the short computation time $t^*$ allows to exclude any effects
caused by sonic waves traveling through the crystal. The velocity of
sound can be estimated as derived in \cite{Hansen}: The equilibrium
distance inside the silicon crystal is approximately
$d_0 = 1.87\cdot 10^{-10}\textrm{m}$. Using the {\em Young modulus}
$Y = d_0\cdot\pdd{H_{\textrm{loc}}}{d_0}(d_0)
= 5.05\cdot 10^{-9}\textstyle\frac{\textrm{kg}\cdot\textrm{m}}{\textrm{s}^2}$
we obtain a velocity of sound of
$c = \sqrt{\frac{Y\cdot d_0}{m_{Si}}} = 4.50\cdot 10^{3}\textstyle\frac{\textrm{m}}{\textrm{s}}$.
The shortest distance of the copper atom to any boundary in the crystal
is $s_{\textrm{left}} = q_{22}-q_1 = 2.01\cdot 10^{-9}\textrm{m}$ in the
original system and
$\tilde s_{\textrm{right}} = q_{50}-q_{22} = 1.90\cdot 10^{-9}\textrm{m}$
in the optimal prediction system. Hence, the minimum time a sonic wave
takes to travel from a boundary to the copper atom is approximately
$t_{\textrm{min}} = \tilde s_{\textrm{right}}/c = 4.22\cdot 10^{-13}\textrm{s}$,
which is longer than the computation time $t^*=4.0\cdot 10^{-13}\textrm{s}$.
In Subsection~\ref{subsec:sonic_waves} we will deal with the case when sonic
waves are present.

\begin{figure}[htb]
\begin{center}
\includegraphics[width=.6\textwidth]{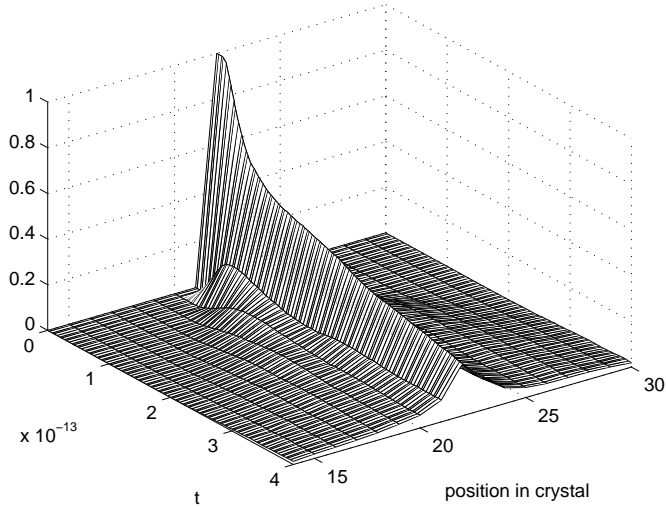}
\caption{Distribution of the copper atom's position}
\label{plot_distr_70_orig.eps}
\end{center}
\end{figure}

Figure~\ref{plot_distr_70_orig.eps} shows the distribution of the position of the
copper atom in the silicon crystal when solving the original system. The analogous
distribution for the optimal prediction system looks indistinguishably similar in
this kind of plot. The process is apparently of a diffusive nature, but the
diffusion parameter changes with time, as already the evolution of the maximum
indicates. The following analysis will confirm this observation.

The time-dependent relative error between the distributions for the original and
the optimal prediction system 
\begin{equation}
e(t) = \frac{\max_{x}|v(x,t)-\tilde v(x,t)|}{\max_{x}|v(x,t)|}
\end{equation}
is less than 3\% up to the time $t=3.0\cdot 10^{-13}\textrm{s}$ and still
less than 9\% over the complete time interval, which is not overwhelmingly
small, but does indicate definite similarities between the two distributions.

\subsection{Diffusion Parameters}
\label{subsec:diffusion_parameters}

We compute the diffusion parameters $\kappa(t)$ for the original and the
optimal prediction system, using the method described in
Subsection~\ref{subsec:model_copper_diffusion}. The parameters
$(p_1,\dots,p_k)$ of the corresponding random walk model are obtained by
Monte-Carlo sampling. We use the experiments for the original system, which
already yielded the distribution shown in Figure~\ref{plot_distr_70_orig.eps}.
In each Monte-Carlo experiment we follow the path of the copper atom and
cluster consecutive hopping events which happen in an interval of
$\Delta t_1=6.0\cdot 10^{-14}\textrm{s}$ to a single one. Prior to this
clustering, fast double hopping events into opposing directions,~i.e. those
happening inside of $\Delta t_2=2.0\cdot 10^{-14}\textrm{s}$, which are
solely due to oscillations of silicon atoms, must be excluded. The times
$\Delta t_1$ and $\Delta t_2$ are suitably chosen for our model problem.

\begin{figure}[htb]
\begin{center}
\begin{minipage}[t]{.48\textwidth}
\includegraphics[width=.98\textwidth]{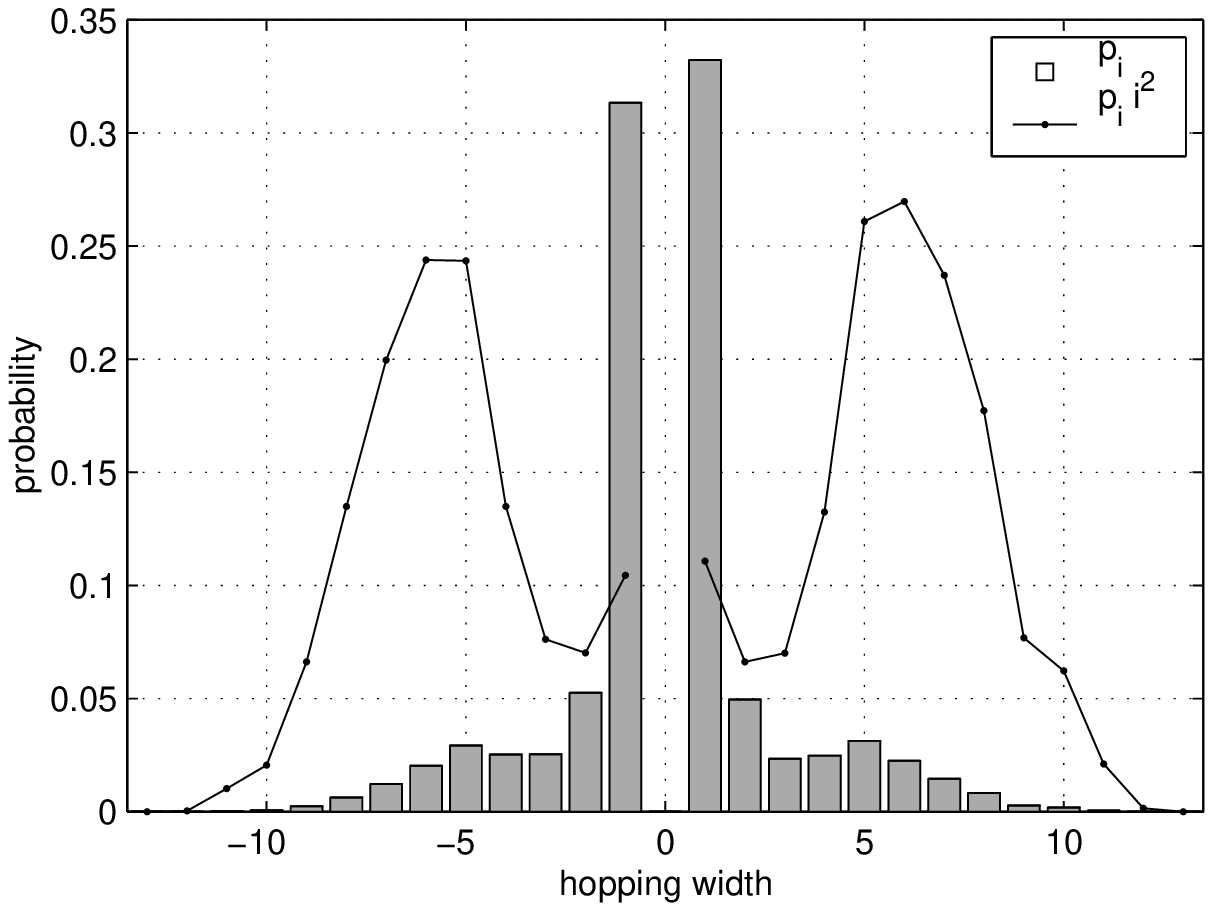}
\caption{Probabilities of hopping over more than a single atom}
\label{plot_hist_hop.eps}
\end{minipage}
\hfill
\begin{minipage}[t]{.48\textwidth}
\includegraphics[width=.98\textwidth]{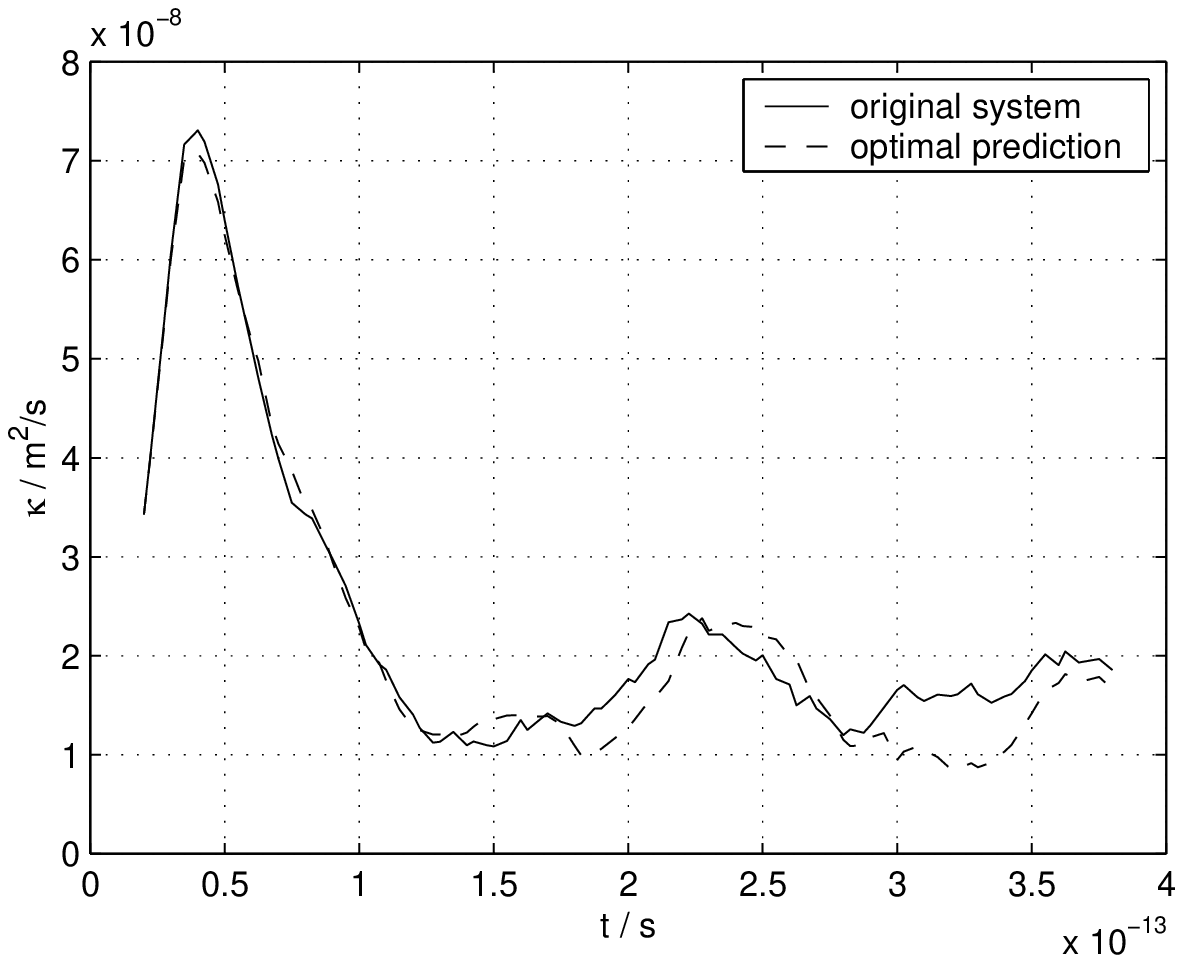}
\caption{Diffusion parameters over time}
\label{plot_graph_kappa.eps}
\end{minipage}
\end{center}
\end{figure}

The results of the above described Monte-Carlo experiments are shown in
Figure~\ref{plot_hist_hop.eps}. The hopping probabilities are given by
the box histogram. Note that due to Monte-Carlo errors and boundary effects
the resulting values are not exactly symmetric. Since we wish to consider
a symmetric random walk, we choose as $p_1,\dots,p_k$ the average values of
the obtained results. Here, only $p_1,\dots,p_{11}$ are greater than zero.
Hence, we choose $k=11$ for the computation of the diffusion parameters.
The curve denotes the values $i^2p_i$ (scaled to fit into the same plot),
which are relevant, since the variance $\sum_{i=1}^k i^2 p_i = \frac{\gamma}{d_0^2}$
is proportional to $\kappa$ (see relation (\ref{relation_kappa_alpha})).

Figure~\ref{plot_graph_kappa.eps} shows the time-dependent diffusion
parameter $\kappa(t)$ for the original system and the optimal prediction
approximation, computed as described in
Subsection~\ref{subsec:model_copper_diffusion}. In both cases we choose
$\Delta t = 2.5\cdot 10^{-14}\textrm{s}$. Important observations and
estimates are:
\begin{itemize}
\item
The diffusion parameters start with a strong peak, and after
$1\cdot 10^{-13}\textrm{s}$ they fluctuate around a fixed value of
$\kappa = 4.4\cdot 10^{-10}\frac{m^2}{s}$. This behavior is too pronounced
to be only due to Monte-Carlo and approximation errors. The initial behavior
is most likely due to the fact that the system does not exactly start in its
thermodynamical equilibrium.
\item
Considering that $\kappa(t)$ shows such pronounced behavior, the two
functions $\kappa(t)$ for the original system, and $\tilde\kappa(t)$ for
the optimal prediction approximation are remarkably close to each other.
Up to the time $t=1.4\cdot 10^{-13}\textrm{s}$ the the distance between
the two curves is quite small. After that time the curves differ more,
but show the same features. After 
$3.0\cdot 10^{-13}\textrm{s}$ the error takes its maximum, which
coincides with the errors observed in the distributions shown in
Figure~\ref{plot_distr_70_orig.eps}.
\item
In order to roughly estimate whether the obtained diffusion parameter is
reasonable, we compare it with a diffusion coefficient measured in a
real material. In \cite{DiffConst} the diffusion constant of copper in a
silicon crystal at a temperature of $T_0 = 1273\textrm{K}$ is given as
$\kappa_0 = 4.4\cdot 10^{-10}\frac{m^2}{s}$. Assuming that the diffusion
constant depends linearly both on temperature and the potential barrier,
as e.g.~with the Einstein formula \cite{Hansen}, we obtain an estimate
$\kappa \approx \kappa_0 \cdot \frac{T}{T_0} \cdot \frac{\Delta E_0}{\Delta E}
= 1.7\cdot 10^{-8}\textstyle\frac{m^2}{s}$ for
$T = 7000\textrm{K}$, using the values $\Delta E_0 = 3\textrm{eV}$ and
$\Delta E = 0.43\textrm{eV}$.
A look at Figure~\ref{plot_graph_kappa.eps} indicates that our
experimentally obtained diffusion parameters are indeed in this region.
\end{itemize}

\subsection{The Number of Hopping Events}
\label{subsec:number_hopping_events}

\begin{figure}[htb]
\begin{center}
\begin{minipage}[t]{.48\textwidth}
\includegraphics[width=.98\textwidth]{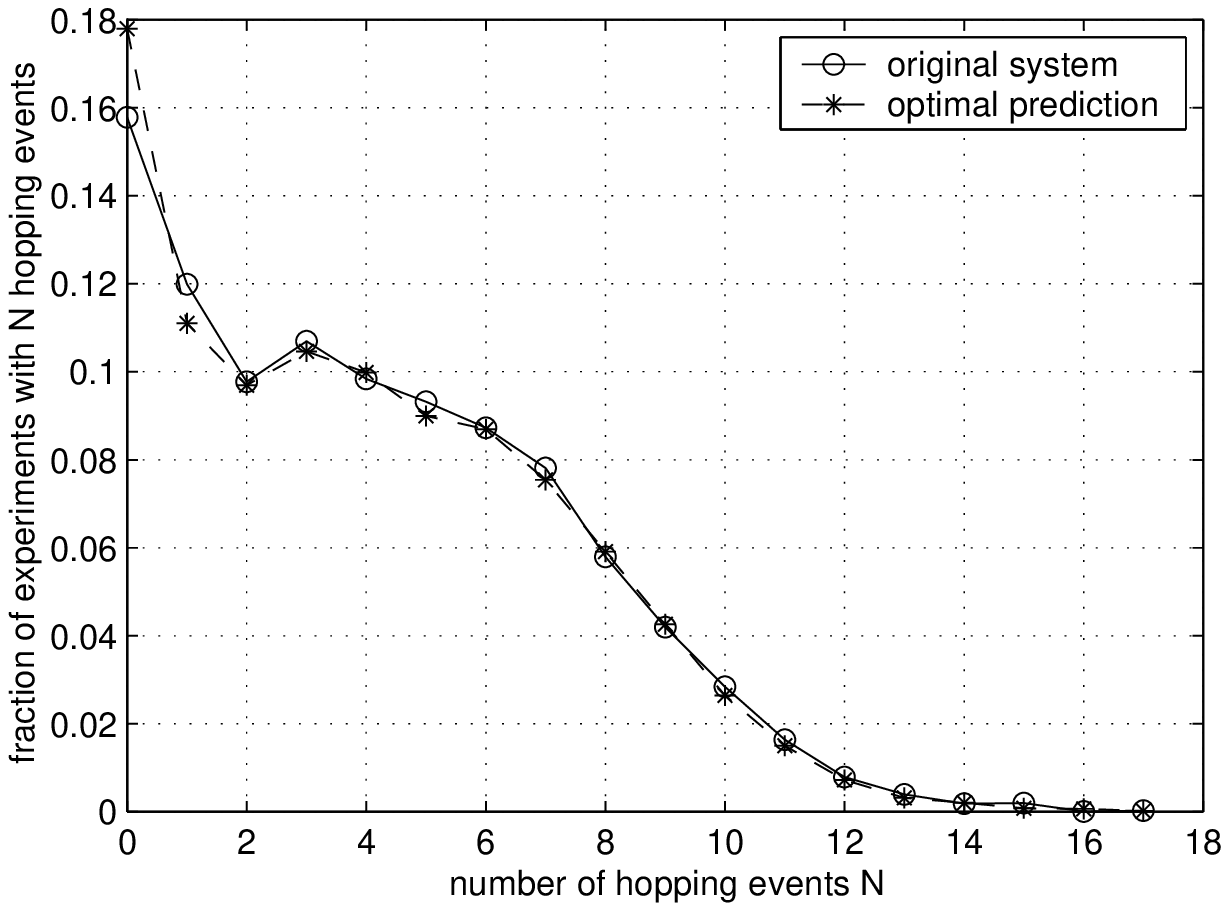}
\caption{Number of hopping events}
\label{plot_graph_histhopp_70.eps}
\end{minipage}
\hfill
\begin{minipage}[t]{.48\textwidth}
\includegraphics[width=.98\textwidth]{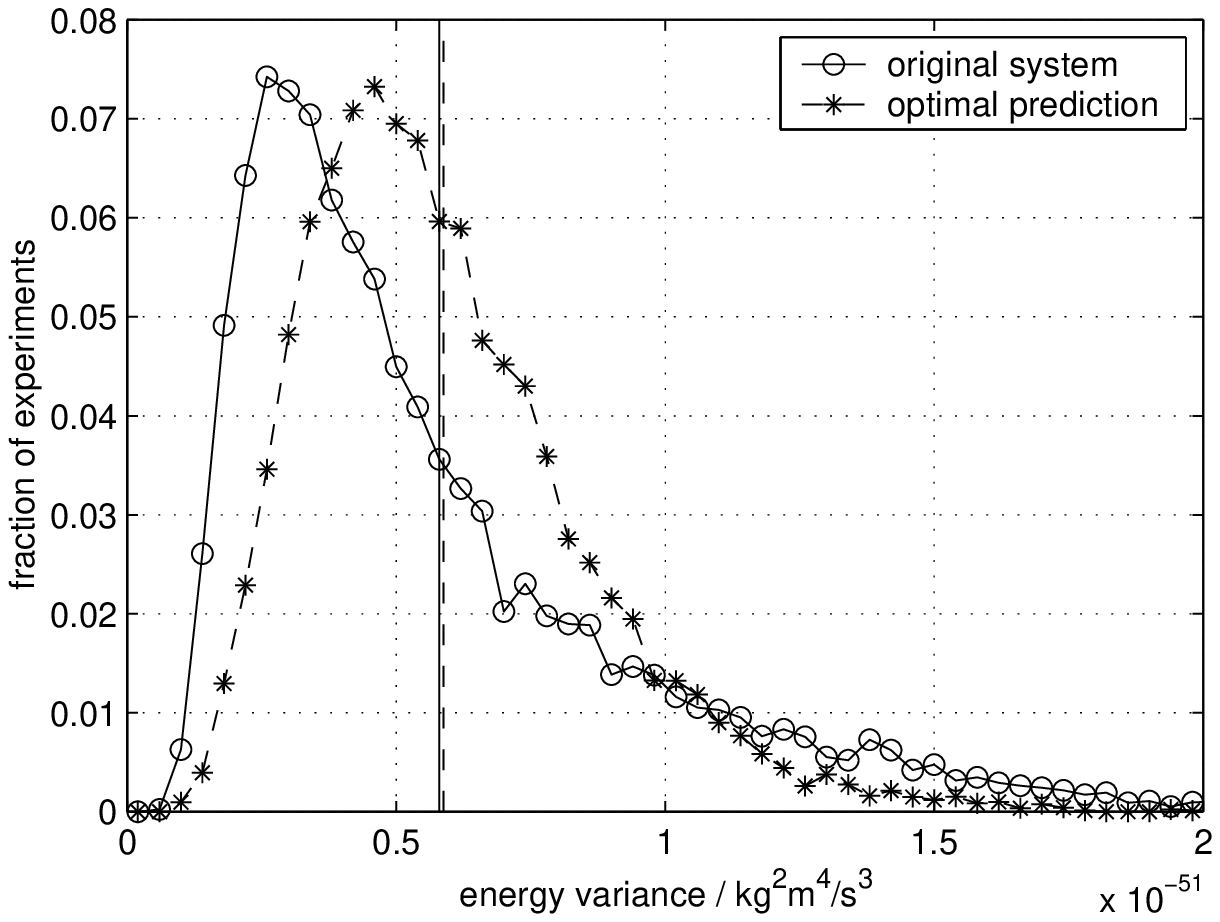}
\caption{Total energy fluctuation}
\label{plot_graph_energyfluct.eps}
\end{minipage}
\end{center}
\end{figure}

Besides diffusion parameters, also the pure number of hopping events which
happen up to a given time $t^*$ should be preserved by optimal prediction.
Figure~\ref{plot_graph_histhopp_70.eps} shows the number of hopping events
for the above computation plotted as histograms. The solid line represents
the original system, and the dashed line stands for the optimal prediction
system. The two graphs differ only for the probabilities of zero and one
hopping event. Apart from that, one can speak of the same hopping behavior.

\subsection{Energy Fluctuations}
\label{subsec:energy_fluctuations}

While the total energy is constant for the original system as well as for
the optimal prediction approximation, the energy of the first $m$ atoms
\begin{equation}
E_{\textrm{left}}(t) = \frac{1}{2}\sum_{i=1}^{m}\frac{p_i^2(t)}{m_i}
+\dsum{i,j=1}{i<j}{m}f_\alpha(q_i(t)-q_j(t))
\label{energy_of_real_atoms_time}
\end{equation}
fluctuates over time. This expression fluctuates also for the optimal
prediction system, since the influence of the virtual atoms is neglected in
the potential energy. The fluctuations in (\ref{energy_of_real_atoms_time})
represent the exchange of energy between atoms, which is a quantity that
should be preserved. Since for optimal prediction we consider the energy of
exactly the real atoms, this test enlightens the exchange of energy between
real and virtual atoms. For each Monte-Carlo experiment we consider the
variance of (\ref{energy_of_real_atoms_time}) over time
\begin{equation}
V = \int_{t=0}^{t^*}\brk{E_{\textrm{left}}(t)-E_{\textrm{left}}(0)}^2\ud{t},
\end{equation}
which measures the impact of fluctuation. Hence, we obtain $N$ values
$V_1,\dots,V_N$ for both the original system and the optimal prediction
approximation.

Figure~\ref{plot_graph_energyfluct.eps} shows the histogram for the variances
$V_i$ for the two systems. The solid line stands for the original system,
and the dashed line corresponds to optimal prediction. The average energy
fluctuations for both systems are denoted by the corresponding vertical lines.
While the average fluctuations are very close for the two systems and the two
distributions look similar in principle, they obviously do not coincide. For
optimal prediction most energy fluctuations are stronger than for the
original system. On the other hand, particularly high energy fluctuations
happen more frequently in the original system. Possible physical reasons for
this behavior could be:
\begin{itemize}
\item
The fact, that in optimal prediction most fluctuations are stronger than in the
original system might be due to the additional fluctuative {\em Langevin} terms,
which appear in higher order optimal prediction \cite{CHK1}.
\item
The high fluctuations in the original system can happen, since around the
$m^\mathrm{th}$ atom energy can be exchanged freely among a whole cluster
of atoms. In the zeroth order approximation to optimal prediction
(\ref{asymptotic_expansion_zeroth_order}), on the other hand, there is no
free energy exchange between the virtual atoms, since they have no momentum,
but instead always follow the potential minimum. Using the first order
asymptotic expansion (\ref{asymptotic_expansion_first_order}) might remedy
this problem.
\end{itemize}
Additionally, at correct temperatures ($T=500\textrm{K}$ instead of
$T=7000\textrm{K}$) the energy fluctuations in optimal prediction might be
much closer to the truth, even for the zeroth order approximation.

\subsection{Sonic Waves}
\label{subsec:sonic_waves}

In the above experiments care was taken to exclude sonic waves, which was
done by keeping the computation time shorter than a wave would take to
travel through half the crystal. If the computation time is quadrupled
and the diffusion distribution analogous to the one shown in
Figure~\ref{plot_distr_70_orig.eps} is computed, one can indeed observe a
small antidiffusive peak every $5\cdot 10^{-13}\textrm{s}$, i.e.~in many
experiments the copper atom is systematically pushed back to its starting
position. The fact that the above time equals approximately the time a sonic
waves takes for traveling from the crystal boundaries to the copper atom
gives rise to the assumption that the observed behavior is indeed due to
sonic waves.

The relevance of sonic waves for optimal prediction can be seen if one
lets a sonic wave run into the boundary between real and virtual atoms.
The wave does not penetrate into the block of virtual atoms, but is being
{\em reflected} by them. Hence, optimal prediction simulates a crystal
continued to infinity only as long as the system is in thermodynamical
equilibrium. Non-equilibrium effects, in particular sonic waves, are not
reproduced correctly by the optimal prediction system in the presented
form. Indeed, in the presence of sonic waves optimal prediction yields
a very different copper diffusion, unlike the experiments where sonic
waves were excluded. Thus the influence of non-equilibrium effects has
to be considered carefully when optimal prediction is applied to other
problems.

\section{Conclusions and Outlook}

In this paper, we applied the method of optimal prediction to a model
problem in the context of molecular dynamics, focusing on diffusion
due to atomic hopping.
Employing the fact that the temperature of the process is low, asymptotic
methods were applied to evaluate the conditional expectations which arise
in optimal prediction. The zeroth order asymptotic expansion was used to
derive a new system of equations, which is formally smaller. Since in
molecular dynamics potentials typically range only over short distances,
the new system could be cut off after a boundary layer of virtual atoms.
This {\em boundary layer condition} acts as if the crystal was continued
to infinity, and yields an obvious computational speed-up for our model
problem. The asymptotic method itself should apply to much more general
cases than the specific model problem here.

In order to investigate whether the thus derived system preserves the
statistical behavior of the original system, various criteria were
introduced and checked by Monte-Carlo experiments. In particular, the
diffusion of a copper atom in the crystal as well as the number of
hopping events turned out to be preserved well. The exchange of energy at
the boundary layer was not preserved that well, but this discrepancy
could be explained. On the other hand, the new system yielded significantly
worse results in the presence of non-equilibrium effects, in particular
sonic waves, since optimal prediction assumes the system to be in
equilibrium.

A natural next step in research on this topic would be to apply the method
to a more complex problem, possibly in three space dimensions and with
focus on further effects than atomic hopping. The basic ideas presented in
this paper should also apply in three space dimensions, but various
aspects will become more problematic. On the other hand, in three space
dimensions sonic waves should dissipate faster, and the fraction of atoms
which can be averaged out would be larger. Additionally, the relevant
statistical quantities should be approximated much better at physically
correct temperatures. An obvious step for deriving a more accurate system
is to employ the first order asymptotic expansion, which was derived in
Section~\ref{sec:op_applied}. Of particular interest in this context is
the question, how to generalize the method of optimal prediction to
problems not in equilibrium.

We have shown that optimal prediction can in principle be applied to
problems in molecular dynamics which take place at comparably low
temperatures and are in equilibrium. The boundary condition version
yielded an obvious speed-up. While the pure speed-up is not overwhelming
by itself, the physical interpretation of the optimal prediction
equations should allow to apply the method in combination with other
methods. Further investigation may improve the derived results.

\section*{Acknowledgements}
We would like to thank
Helmut Neunzert, Thomas G\"otz and Peter Klein for useful advice and
encouragement and Alexandre J. Chorin, Mi\-chael Junk and Jochen Vo\ss{}
for helpful discussions and comments.



\begin{thebibliography}{99}

\bibitem{BCC}
J. Bell, A.J. Chorin and W. Crutchfield,
{\em Stochastic optimal prediction with application to averaged Euler equations},
Proc. 7th Nat. Conf. Comput. Fluid Mech., C.A. Lin (ed), Pingtung, Taiwan,
pp. 1-13, 2000.

\bibitem{CageCorr}
B.J. Berne, J.D. Gezelter, E. Rabani,
{\em Calculating the hopping rate for self-diffusion on rough potential energy
surfaces: Cage correlations},
J. Chem. Phys. 107, No. 17, pp. 6867-6876, 1997.
 
\bibitem{DiffConst}
L.J. Chen, S.L. Cheng, H.H. Lin,
{\em The failure mechanisms and phase formation for Ni, Co and Cu contacts
on ion implanted (001)Si under high current stress},
Nucl. Instr. and Meth.in Phys. Res. B 169, pp. 161-165, 2000.

\bibitem{CHK1}
A.J. Chorin, O. Hald and R. Kupferman,
{\em Optimal prediction and the Mori-Zwanzig representation of irreversible
processes},
Proc. Nat. Acad. Sc. USA, 97, pp. 2968-2973, 2000.

\bibitem{CHK2}
A.J. Chorin, O. Hald and R. Kupferman,
{\em Non-markovian optimal prediction},
Monte Carlo Methods and Applications, Vol. 7, No. 1-2, pp. 99-109, 2001.

\bibitem{CHK3}
A.J. Chorin, O. Hald and R. Kupferman,
{\em Optimal prediction with memory},
Physica D 166, No. 3-4, pp. 239-257, 2002.

\bibitem{CKK1}
A.J. Chorin, A. Kast and R. Kupferman,
{\em Optimal prediction of underresolved dynamics},
Proc. Nat. Acad. Sc. USA, 95, pp. 4094-4098, 1998.

\bibitem{CKK2}
A.J. Chorin, A. Kast and R. Kupferman,
{\em Unresolved computation and optimal prediction},
Comm. Pure Appl. Math., 52, pp. 1231-1254, 1999.

\bibitem{CKK3}
A.J. Chorin, A. Kast and R. Kupferman,
{\em On the prediction of large-scale dynamics using unresolved computations},
Contemp. Math., 238, pp. 53-75, 1999.

\bibitem{ChorinNotes}
A.J. Chorin,
{\em Probability, mechanics, and irreversibility},
Lecture notes, UC Berkeley Math. Dept., 2000.

\bibitem{CKL}
A.J. Chorin, R. Kupferman and D. Levy,
{\em Optimal prediction for Hamiltonian partial differential equations},
J. Comput. Phys., 162, pp. 267-297, 2000.

\bibitem{Devroye}
L. Devroye,
{\em Non-uniform random variate generation},
Springer, 1986.

\bibitem{ProblemSimulationMD}
T. Frauenheim, H. Hensel, P. Klein, H.M. Urbassek,
{\em Comparison of classical and tight-binding molecular dynamics for
silicon growth},
Phys. Rev. B 53, pp. 16497-16503, 1996.

\bibitem{FMM}
L. Greengard, V. Rokhlin,
{\em A fast algorithm for particle simulations},
J. Comp. Physics, 73, pp. 325-348, 1987.

\bibitem{HK1}
O. Hald and R. Kupferman,
{\em Convergence of optimal prediction for nonlinear Hamiltonian systems},
SIAM J. Num. Anal. 39, pp. 983-1000, 2001.

\bibitem{Hammersley}
J.M. Hammersley, D.C. Handscomb,
{\em Monte Carlo methods},
Chapman and Hall, 1979.

\bibitem{Hansen}
J. P. Hansen, I. R. McDonald,
{\em Theory of simple liquids},
Academic Press, London, 1976.

\bibitem{ProblemExperiment}
S. Hasegawa, S. Ino, Z.H. Zhang,
{\em Epitaxial growth of Cu onto Si(111) surfaces at low temperatures},
Surface Science 415, pp. 363-375, 1998.

\bibitem{K1}
A. Kast,
{\em Optimal prediction of stiff oscillatory mechanics},
Proc. Natl. Acad. Sci. USA 97, No.12, pp. 6253-6257, 2000.

\bibitem{Annealing}
S. Kirkpatrick, C.D. Gelatt Jr, M.P. Vecchi,
{\em Optimization by Simulated Annealing},
Science, V. 220, No. 4598, pp. 671-680, 1983.

\bibitem{Krengel}
U. Krengel,
{\em Einf\"uhrung in die Wahrscheinlichkeitstheorie und Statistik},
Vieweg, 1991.

\bibitem{Lawler}
G.F. Lawler,
{\em Introduction to stochastic processes},
Chapman and Hall, London, 1995.

\bibitem{FicksLaw}
M.H. Lee,
{\em Fick's law, Green-Kubo formula, and Heisenberg's equation of motion},
Physical Review Letters, Col. 85, No. 12, pp. 2422-2425, 2000.

\bibitem{Murray}
J.D. Murray,
{\em Asymptotic analysis},
Springer, 1984.

\bibitem{Olver}
F.W.J. Olver,
{\em Asymptotics and special functions},
Academic Press, 1974.

\bibitem{DA}
B. Seibold,
{\em Optimal prediction in molecular dynamics},
Diploma Thesis, University of Kaiserslautern, 2003.

\bibitem{Constants}
B.M. Smirnov, A.S. Yatsenko,
{\em Properties of dimers},
Phys. Usp. 39, No. 3, pp. 211-230, 1996.

\end{thebibliography}
\end{document}